%% file: kosenko.tex
\documentclass[]{aa}
\input{defs}

\usepackage{txfonts}
\usepackage{natbib}
\bibpunct{(}{)}{;}{a}{}{,}

\begin{document}

\title{XMM-Newton X-ray spectra of the SNR~0509-67.5: data and models}
\titlerunning{The X-ray Spectra Of The SNR 0509-67.5}

\author{D. Kosenko\inst{1,3} \and J. Vink\inst{1,2} \and S. Blinnikov\inst{4,3} \and A. Rasmussen\inst{5}}
   \institute{
Astronomical Institute Utrecht,  University Utrecht, P.O. Box 80000, 3508TA Utrecht, The Netherlands \\
              \email{D.Kosenko@astro.uu.nl}
         \and
SRON, Utrecht, The Netherlands
         \and
Sternberg Astronomical Institute, Russia
         \and
Institute for Theoretical and Experimental Physics, 
117218, Moscow, Russia
         \and
Stanford Linear Accelerator Center, Menlo Park, CA, USA
             }

\authorrunning{Kosenko et al.}
\date{Received ...; accepted ... }

\abstract{}{
We report on X-ray observations of the supernova remnant 0509-67.5 in the Large
Magellanic Cloud with XMM-Newton X-ray observatory. We use the 
imaging spectroscopy (EPIC) and Reflective Grating Spectrometer (RGS) data
to investigate properties of the remnant and its environment.
}{
The X-ray spectra were analyzed with SPEX software package. 
In addition to this we performed a numerical hydrodynamic simulation of the remnant.
}{
The EPIC data show prominent Fe~K line emission, but the deduced overall amount of
iron in the shocked ejecta is low. The data also show that the remnant has
an asymmetric ejecta structure: the bright southwest region of the remnant
shows an overabundance of metals. The analysis of the RGS spectrum shows
that the remnant has a high lines velocity broadening of ~5000 km/s. We
found a hydrodynamical model for the remnant with basic hydrodynamical and
spectral parameters similar to the observed ones.
}{
The data analysis show that the reverse shock just recently reached iron
layers of the ejecta. The brightness enhancement in the southwest region could be a
sign of an asymmetric explosion or it could be the result of a density
enhancement of the interstellar medium. We constructed numerical models
which are in good agreement with the observations, with circumstellar
density of
$3\times10^{-25}\;\mt{g/cm}^3$, age of $\sim 400$ years, velocities of 
$\sim 5000$ km/s and an electron to ion temperature ratio of $10^{-2}$.
}
\keywords{X-rays: individuals: SNR0509-67.5 --- ISM: individuals objects: SNR0509-67.5 --- 
ISM: supernova remnants --- method: data analysis --- hydrodynamics}

\maketitle

\section{Introduction} 

The X-ray emission from young supernova remnants (SNRs) provide a vital source
of information about the properties of shocked, rarefied plasma, properties
of the ambient medium around the remnants, and also about supernova
explosion models. 
With the current generation of X-ray telescopes, XMM-Newton, Chandra,
and Suzaku, one can combine now the imaging and spectral information
to study the extended emission from SNRs. Moreover, in the case of
Chandra and XMM-Newton, dispersive spectrometers 
allows one to obtain high resolution spectra, providing more constraints
on the plasma parameters, and on the dynamics of the plasma, through
Doppler shifts and broadening. The XMM-Newton Reflection Grating 
Spectrometers (RGS) have the advantage that the spectral quality is less
detoriated by the spatial extent of the object, at least for sources
with an angular extent of $\lesssim 1$~\arcmin. 

For that reason the Large Magellanic Cloud (LMC) and the Small Magellanic Cloud
(SMC) SNRs are particular well suited, since the relative proximity of the
LMC means that many SNRs are bright enough to obtain high signal to noise
spectra with both the CCD and grating spectrometers. On the other hand, the
distance of the LMC, 50~kpc, is such that young SNRs have an extent less
than 1~\arcmin, making them excellent targets for the XMM-Newton RGS. The
XMM-Newton CCDs, with a resolution of $\sim 6$~\arcsec, still allows one to
study spatial variations, although Chandra obviously provides more spatial
details. In addition, studying SMC/LMC remnants has the advantage that the
interstellar absorption is on average lower than for most of the Galactic
remnants, and their distance is known.

XMM-Newton observations of several LMC/SMC remnants have been 
published: N132D \citep{behar}, 1E~0102.2-7219 \citep{rasmussen,sasaki},
B0540-69.3 \citep{vanderheyden_b}, N103B \citep{vanderheyden_n}, and several
SMC remnants \citep{vanderheyden_smc}.

Here we report on the X-ray spectrum of the SNR~0509-67.5 in the Large
Magellanic Cloud. This object was already examined by several authors, e.g.,
\citet{warren,jvink,badenes08,rest08,parviz_uv,rest,badenes07,hendr}. The
remnant has a spherical shape and is somewhat similar in age and size to
Tycho's SNR, which is also a Type Ia SNR. SNR~0509-67.5 has an angular
diameter of 25\arcsec, corresponding to the radius of 3.6 pc \citep{warren}. 
A recent analysis by \citet{rest}, based on light echo detection,
and
\citet{badenes08}, based on modeling of the X-ray observation, suggest the
age of the remnant to be $\sim400$ years.
Estimates by \citet{parviz_uv} gives the values of $295 - 585$ years,
consistent with the rough estimate $\sim 500$ years by \citet{jvink}, based on
a preliminary analysis of the XMM-Newton data. 

A detailed analysis of the X-ray emission of SNR~0509-67.5 was made by
\citet{warren} and based on Chandra X-ray data. Their study revealed that
the remnant is rich in silicon, sulfur and iron, the bulk of the continuum
emission has a non-thermal origin, and an estimate the circumstellar medium
(CSM) density is of $\sim 0.05\;\mbox{cm}^{-3}$. The preferred model for the
explosion which produced the remnant --- is delayed detonation.

Recently, \citet{badenes08} reported their study of the remnant's
X-ray spectrum observed by Chandra and XMM-Newton observations. They
performed hydrodynamical and X-ray spectral calculations of the remnant and
concluded that it was an energetic, delayed detonation explosion of
$1.4\times10^{51}$ ergs with a nickel mass of $0.97M_\odot$. In their
numerical simulations the circumstellar density was set to $n_\mt{CSM} =
0.4\,\mt{cm}^{-3}$ and the ratio of electron to ion temperatures to $0.02$.

\begin{figure}
\begin{center}%
        \includegraphics[width=0.29\hsize]{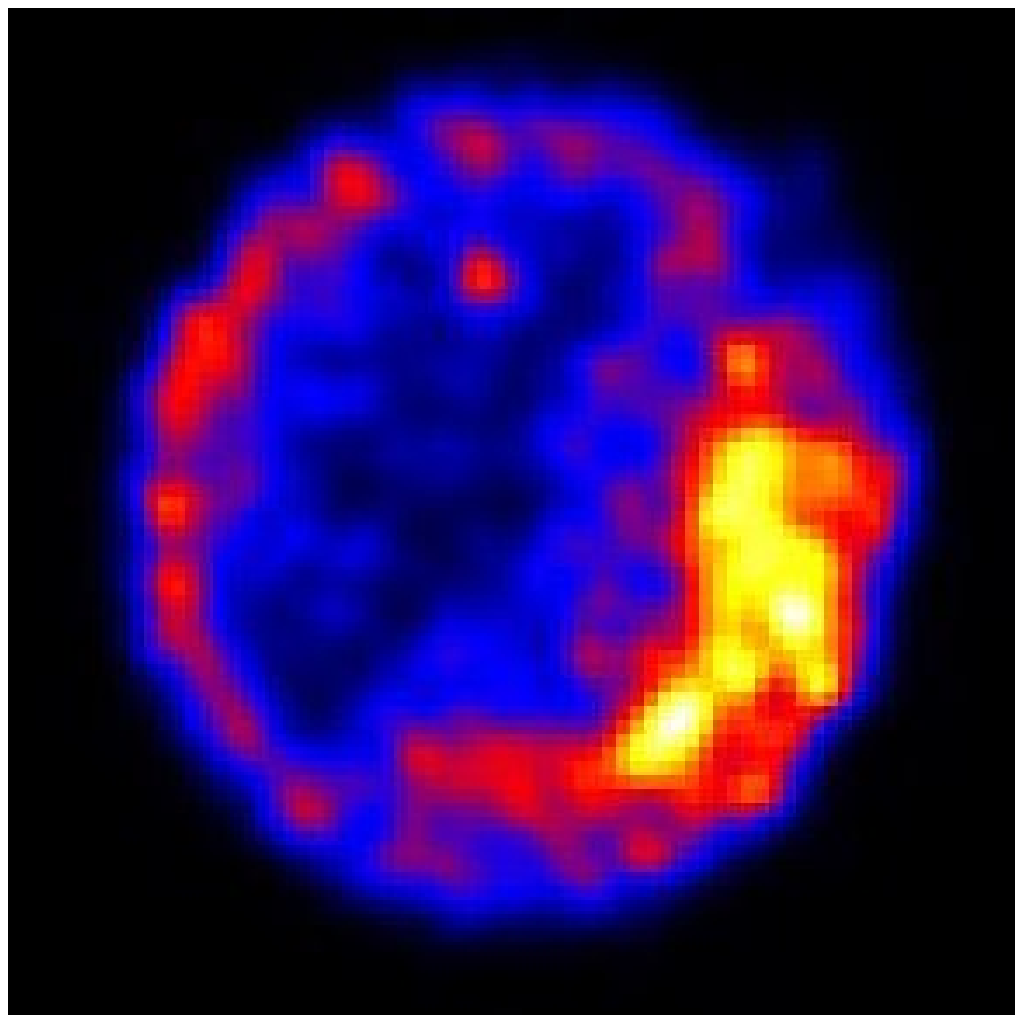}
        \includegraphics[width=0.29\hsize]{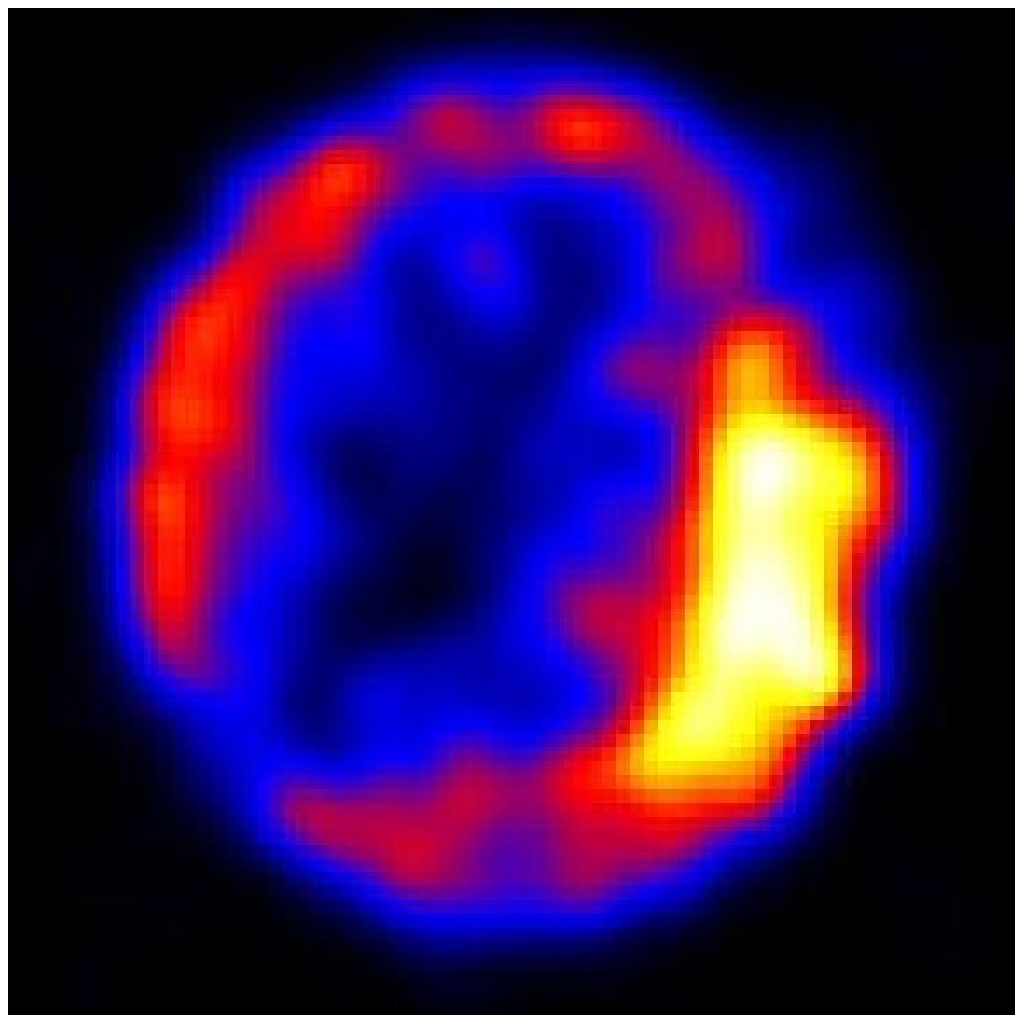}
        \includegraphics[width=0.29\hsize]{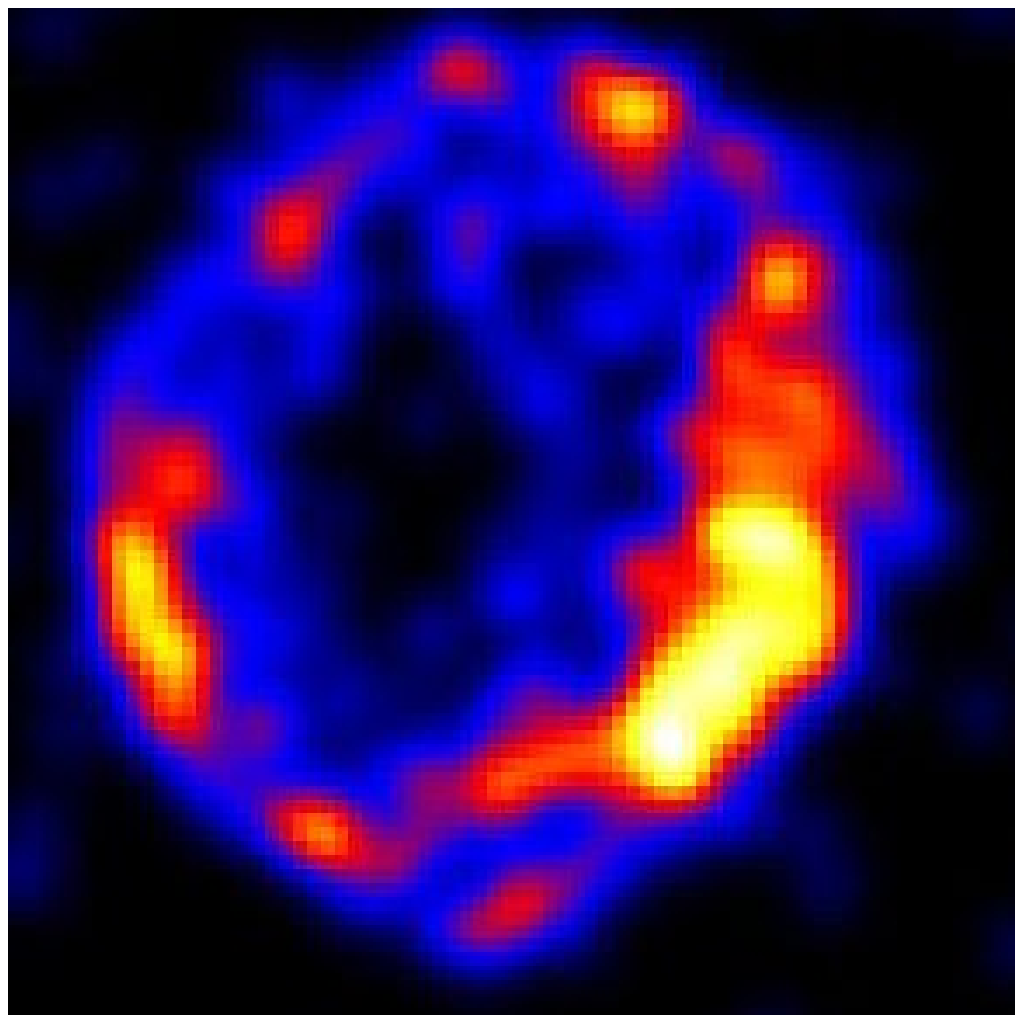}
        \caption{Chandra images of the SNR~0509-67.5 in different X-ray ranges. From left to right: 0.45-1.75 keV, 1.8-2.5 keV, 2.5-6.0
        keV \citep{warren}.}
        \label{chandra}
\end{center}\end{figure}

For illustrative purposes we show in \rfig{chandra} some of the Chandra
X-ray images of the remnant in three different bands. The overall structure
of the SNR shows a large scale inhomogeneity in the shell. The Southwest
(SW) part of the remnant is somewhat brighter compared to the average
brightness.

In the current study we performed an analysis of the remnant employing
XMM-Newton X-ray observations and {\sc spex} fitting software \citep{spex}.
We also performed hydrodynamical simulations of the SNR~0509-67.5, employing
the hydrocode {\sc supremna} \citep{sorokina} to compare
numerical predictions for the remnant with the observations.

The paper is organized as follows. First we describe briefly the XMM-Newton
data in $\S$\ref{data}, then we discuss the X-ray spectrum in
$\S$\ref{spectra}. Numerical models for the SNR are presented in
$\S$\ref{models}. We discuss our results and outline them in
$\S$\ref{discs}.

\section{The XMM-Newton data, reduction and analysis} \label{data} 

SNR~0509-67.5 was observed with the XMM-Newton X-ray Observatory (obs
ID~0111130201) for 35.9 ks on July 4, 2000. The observatory contains three X-ray
telescopes. Behind all of them are CCD detectors, collectively called the
European Photon Imaging Camera (EPIC). Two of them are MOS type CCD
detectors called MOS1 and MOS2 \citep{turner}, the other is of the pn-CCD
type \citep{struder}. The telescopes equipped with MOS CCD detectors in the
focal plane contain also the Reflective Grating Spectrometers (RGS),
consisting of the Reflective Grating Arrays, dispersing about 50\% of the
X-rays to two CCD arrays \citep{denherder}. XMM-Newton observations
therefore provide simultaneous data from all five X-ray instruments: the two
MOS-CCDs, the pn-CCD, and the two RGS instruments.

In this paper we concentrate on the data obtained with the MOS detectors and
the RGS. Although the EPIC-MOS have a somewhat lower sensitivity than the
EPIC-pn instrument, they have better spectral resolution, which is important
for line rich sources, such as SNRs.

Note that~\cite{badenes08} found an inconsistency between EPIC
MOS1,2 and PN spectra. They investigated Si~$K_\alpha$ line centroid
location in the datasets of these devices and found out that the line
centroids of MOS and PN detectors are shifted with respect to each other.
The authors rejected MOS1 and MOS2 data and concentrated only on the Chandra
and EPIC-pn spectra. We confirm this inconsistency: analysis of the
Si~$K_\alpha$ centroid of MOS1,2 and PN data reveals that it is located at
$1.8503\pm0.0012$ keV and
$1.8298\pm0.0015$ keV respectively (the errors are $1\sigma\; \chi^2$). For
the Fe~$K_\alpha$ line we found $6.495\pm0.048$~keV and
$6.442\pm0.025$~keV. The discrepancies in the lines locations are about 1\%.
Nevertheless, there are indications, that PN instrument has small gain
problems and sometimes a shifting of the PN energy grid is
required~\citep[private communications with N.Werner;][]{werner06,plaa04}.
Moreover, in our study, we were unable to find a fitting model with sensible
parameters of the emitting plasma for the PN data, meanwhile a spectral model
for the MOS data gives reasonable and satisfactory plasma parameters.

The EPIC data was slightly affected by background flaring (``soft
protons''). As a result we cut out 3.4 ks of the total observation. The MOS1
observation was done in small window mode, whereas the MOS2 data were made
in full frame mode. For MOS1, therefore, we could not select a region to
extract background spectra. For the MOS2 this was possible, but we found
that the background correction was too small, and the account for it did not
change the fitted parameters. For that reason we did not include background
subtraction for our final analysis.

The RGS is a slitless spectrometer. For an extended source this means that
the spectrum is smeared by the image of the source itself.
For SMC/LMC remnants the smearing is modest, but present, and it gives rise
to a change in the line spread function. For our analysis we incorporated
this effect into the response matrix by convoluting the standard (point source)
response matrix with the emissivity profile of the SNR, as obtained from
archival Chandra observations. This procedure was also for the RGS data of
SN 1006 \citep{vink03}. As we shall show later, however, in the case of
0509-67.5 the Doppler line broadening is much larger than broadening due
to the spatial extent of the remnant.

Apart from adapting the RGS response matrix, all reduction for both MOS and
RGS data  was done using the standard XMM-Newton software package SAS 
version~7.1.0.

\section{Spectral models} 
\label{spectra} 
The X-ray spectra of the SNR were fitted with a non-equilibrium ionization
(NEI) model, which is a part of the {\sc spex} fitting
package~\citep{spex}. 
The package does not include a plane shock model (which takes into account
temperature and ionization timescale gradients), but it contains the most
complete and up-to-date set of atomic data. The use of a single ionization
timescale NEI model is justified in some cases, when one needs to get rough
estimates of basic SNR properties \citep[e.g.][]{chenai04}. The
emission measure $n_Hn_eV$ in such a model can be used to estimate
the circumstellar medium density. Since continuum emission stems mostly from
the shocked CSM, the value of $n_H$ can give us directly
an estimate of the unshocked environment density (depending on the assumed
equation of state).

Combined fitting of the spectra from RGS, EPIC: MOS1,
MOS2 devices is presented in \rfig{spex_sp}.

\subsection{EPIC data}

\subsubsection{Spectra parameters}

In order to adequately fit the spectra we needed at least two thermal
components: one NEI component is for the bulk of the X-ray line emission, and
another one is to fit the Fe-K emission around 6.5~keV (see section \ref{fek} for
details on this component). We also investigated a model with an additional
power law continuum, since~\citet{warren} reported indirect evidence for
non-thermal continuum, probably synchrotron emission from $>$ TeV electrons.
The plasma parameters for the two models are indicated in Table~\ref{pars}
as (th) and (th+pow). In addition, we extracted also the spectrum from the
enhanced SW region of the remnant. In general, the parameters are in
agreement with the the ones obtained by \citet{warren}. The enhanced SW
region shows higher ionization timescale, probably reflecting a higher
density.

Using the value of the emission measure, we can estimate the density of the
CSM. Hydrogen emission measure (Table~\ref{pars}, MOS+RGS) is $\mt{EM} =
n_en_H\,V = 1.5\;\times 10^{58}\;\mt{cm}^{-3}$ and the remnant's size is $R
= 3.6\;\mt{pc} \simeq 10^{19}\;\mt{cm}$. Assuming that the emitting shell
has a width of $\Delta R = R/12$, so that the volume
$V_{X} = V_{tot}/4 = \pi\,R^3/3 \simeq 10^{57}\;\mt{cm}^3$, we obtain
$n_en_H \simeq 15\;\mt{cm}^{-6}$. Furthermore, assuming that all the matter
is ionized $n_H \simeq 4\;\mt{cm}^{-3}$, we obtain that CSM number density
for the remnant should be $n_\mt{CSM} \simeq 1\;\mt{cm}^{-3}$ for the
neutral circumstellar environment, which appears to be the case for
SNR~0509-67.5 \citep{parviz_uv}. Here we assumed shock compression ratio to
be $4$ for non-relativistic matter with adiabatic index $\gamma = 5/3$.
However, cosmic ray (CR) acceleration may play an important role in SNR
dynamics. For a cosmic ray dominated shock the adiabatic approach that of a
relativistic gas ($\gamma = 4/3$) gives rise to a compression factor of $7$.
Cosmic ray escape may even increase the compression ratio beyond $7$
\citep{berezhko99}. Tycho's SNR provides some observational evidence for
high compression factors \citep{warren05}. Therefore, in the case of
0509-67.5, the pre-shock density could be as low as $n_\mt{CSM} \lesssim
0.6\;\mt{cm}^{-3}$.

Also, two additional sources of uncertainty concerning the pre-shock
density are the presence or absence of a non-thermal continuum, and 
chemical composition. In particular, our above estimate assumes that most of
the continuum is bremsstrahlung from hydrogen-electron collisions, whereas
it is quite likely that parts of the remnant consist of pure metal plasmas,
in which case bremsstrahlung comes from metal ions and the electrons they
provide, in general this lowers the density estimate. Therefore, our
estimate of $n_{CSM} \simeq 1$~cm$^{-3}$, should be considered as an upper limit.

Abundances of chemical elements for the entire SNR (thermal and nonthermal
continuum) and enhanced SW part for the simple NEI model are presented in
\rfig{abs}. Abundances for some theoretical explosion models \citep{iwamoto,
woosley} are also shown. The accounting for the nonthermal
continuum does not create a considerable difference in the abundances set,
but naturally it makes the remnant more metal-rich. One can see from
the figure that the SW region is more metal abundant compared to overall
remnant values (2-3 times higher amount of metals).

The {\sc spex} fitting shows no evidence for interstellar absorption, which
means that the value of $N_H$ is low for this remnant. The column density
value found by \citet{warren} is $N_H \simeq (5-9)\times
10^{20}\;\mt{cm}^{-2}$. Which is, indeed, the lowest value among all
known LMC supernovae remnants \citep{LMC_hughes}. This could suggest that
the remnant is located on the near side of the LMC.

The power law model for the continuum gives a value for the nonthermal flux
of $F^X_{\mathrm 1 keV} = (2.3 \pm 0.3)\times
10^{-4}\;\mathrm{ph/s/cm^2/keV}$ with power index of $\Gamma = 3.5 \pm 0.1$,
which is steeper than expected for this kind of remnants, e.g. Tycho
SNR \citep[c.f. Tycho's SNR or Cas A][]{gtycho,vinklaming}. 
This value is close to the one obtained by~\citet{warren}: $3.25\pm 0.18$.
The extrapolation of the non-thermal radio flux \citep{hendr} to the X-ray
range yields $F_{\mathrm 1 keV} = 1.4\times
10^{-2}\;\mathrm{ph/s/cm^2/keV}$. Since the X-ray synchrotron spectrum is
likely to be affected by synchrotron losses, this flux should be considered
as an upper limit.

\FIG spex_sp mos_rgs_syn [width=0.7\hsize,angle=-90] 
MOS1, MOS2 and RGS data fitted with the NEI model, the Fe~K line is fitted separately: 
a NEI component with $kT = 3.5$ keV and $n_et = 10^{9}\;\mt{s/cm}^3$.
Vertical lines show the location of three Gaussians, which substitute missing 
in {\sc spex} model Fe~L lines (the energies are 1.102, 1.125 and 1.226 keV). 

\FIG abs MOS_reg_abs [width=\hsize,angle=0] Abundances of the entire remnant
 and its SW region, obtained in fitting of the EPIC (MOS1 and MOS2) spectrum. 
 Abundances for the classical W7 \citep{iwamoto} and for the
 delayed-detonation model c050403m 
 \citep{woosley} are also shown. Theoretical abundances are taken from
 numerical simulations (described below) include swept up ($\sim
 0.7\,M_\odot$) ejecta and shocked LMC circumstellar medium. All abundances
 are solar normalized. Errors are $1\,\sigma$ rms. The silicon abundance
 (with respect to solar) for the entire SNR is 5.0 and for the SW part of it
 is 15.7.

\begin{table*}
\caption{Parameters for the SNR~0509-67.5 (last column contains data for the southwest region), 
EPIC (MOS1,2) and RGS data. Errors are 1 sigma ($\chi^2$-distribution).
th --- basic (single ionization timescale) NEI model, th+pow --- NEI with power-law continuum.}
\begin{center}
\begin{tabular}{llllll}
\hline \\  Parameter  &      
th, MOS+RGS & th, MOS & th+pow, MOS+RGS &  th+pow, MOS & SW, th, MOS  \\
\hline \\ $n_en_H\,V\;\times 10^{58},\;\mt{cm}^{-3}$ & 
$1.15_{-0.12}^{+0.12}$ & $0.98^{+0.10}_{-0.10}$   & 
$0.55^{+0.15}_{-0.16}$ & $0.95^{+0.11}_{-0.10}$   &
$0.07_{-0.04}^{+0.05}$ \\
\hline \\ $kT$, keV                                & 
$4.01_{-0.18}^{+0.23}$ & $5.02^{+0.33}_{-0.28}$    &
$4.55^{+0.22}_{-0.20}$ & $4.98^{+0.30}_{-0.30}$     &
$4.59_{-0.45}^{+0.54}$ \\
\hline \\ $n_et \times 10^{10},\;\mt{s/cm}^3$        & 
$1.41_{-0.03}^{+0.03}$ & $1.62^{+0.05}_{-0.05}$       &
$1.63^{+0.04}_{-0.04}$ & $1.60^{+0.05}_{-0.06}$    &
$1.52_{-0.08}^{+0.08}$ \\
\hline \\ $\sigma_v, km/s$                            & 
$6030 \pm 170$ & $5700\pm 220$ & $5050\pm 180$ & $5350\pm 240$ & $6100\pm 500$\\
\hline \\ $\chi^2/d.o.f.$                            & 
$2.61$ & $1.42$ & $2.29$ & $1.37$ & $1.66$ \\
\hline
\end{tabular}
\end{center}
\label{pars}
\end{table*}

\subsubsection{Fe-K emission}
\label{fek}
Despite of the low amount of iron in the shocked ejecta deduced
with single ionization timescale NEI model, the spectrum has a
pronounced Fe~K feature. Its centroid is located at $6.50\pm 0.05$~keV,
which may correspond to the ions from {\sc Fe~VII} \citep{kaasmewe} with
$n_et \simeq 10^9\;
\mathrm{s/cm}^3$ up to {\sc Fe~XVII} with $n_et \lesssim 10^{10}\;
\mathrm{s/cm}^3$.  We fitted the Fe~K line separately with
the temperature $kT=3.5$ keV and ionization timescale $n_et=10^{9}\;
\mathrm{s/cm}^3$. 
The ionization time is uncertain, but it is impossible to fit the Fe~K emission
with a component with $n_et > 2\times 10^{9}\mathrm{s/cm}^3$, since then
the spectrum would have exhibited much more Fe~L emission than observed.
Given the low $n_et$ it is likely that the Fe~K emission is associated with
pure Fe, recently shocked by the reverse shock.

The separate fitting of the Fe~K feature allows us to make a crude estimate
of the amount of swept up iron layer of the supernova ejecta. Assuming that
the pressure and temperature do not change drastically throughout the shell,
and thus the electron density $n_e$ remains approximately constant, we can
employ the following relation $n_\mt{Fe}V_\mt{Fe} \sim
\left(EM_\mt{Fe}/EM_X\right)(n_\mt{H}V_\mt{X})$.

The value of the emission measure of the pure
iron NEI component, $EM_\mt{Fe} = n_en_\mt{Fe}V_\mt{Fe}
\simeq 4\times10^{54}\;\mt{cm}^{-3}$, basic NEI component, $EM_\mt{X} =
n_en_\mt{H}V_\mt{X} \simeq 10^{57}\;\mt{cm}^{-3}$, and also the density 
$n_\mt{H} \simeq 4\;\mt{cm}^{-3}$ and the volume $V_\mt{X} \simeq
10^{57}\;\mt{cm}^3$ of the emitting shell, yield us the estimate of the 
swept up iron in the remnant of $M_\mt{Fe} = 56\,m_U\,n_\mt{Fe}V_\mt{Fe} \sim
0.05\,M_\odot$ (with $m_U = 1.66\times 10^{-24}$ g, $M_\odot = 2\times
10^{33}$ g). This is less than 10\% of the iron amount in a typical Ia SN
model \citep{nomoto,iwamoto}.

\subsubsection{The line emission deficit around 1.2~keV}

The EPIC (MOS1,2) models do not fit an excess of emission around 1.2~keV.
This feature is also present in the EPIC-pn spectra, and is weakly visible
in the RGS spectra. Therefore, this excess is likely to be real and not
caused by calibration errors or background features. It is possible to fit
this region of the spectrum, but only if we allow for unrealistic
overabundances of either Na (approximately 5 times of Si) or Ni (4 times of
iron).

In our view the possible cause is the uncertainty in the atomic data base
of SPEX concerning the Fe-L emission. The Fe-L emission of SNR 0509-67.5
is dominated by Fe XVII emission. So it is likely that the excess
is due to incompleteness of the atomic database concerning Fe XVII,
in particular due to line emission from the $n=4$ to $n=2$ level.
This is corroborated by the fact that the series limit of Fe XVII is
at 1.26~keV, which also roughly marks the end of the emission excess.
Some  $n=4 \rightarrow 2$ lines are present in the SPEX code, but their
emission strengths may be underestimated. This may not be surprising,
since the code has been tested primarily on collisional equilibrium
sources, whereas in NEI $n=4 \rightarrow 2$ may be enhanced due to
non-equilibrium effects, such as inner shell ionization of Fe XVI.\footnote{
We discussed this issue with Dr.~Kaastra, one of the authors of the SPEX code, 
who agrees that this may explain the lack of a good fit around 1.2~keV.}

\subsection{RGS data}

The high resolution RGS spectra allowed us to evaluate the velocity
broadening of the X-ray emission lines and make an independent estimate of
the CSM density. Thus, from the spectrum in energy range of $0.3-2.5$ keV we
find $\sigma_v = 4900 \pm 420$ km/s (which is the preferred value since it
is based on the RGS only). This gives us an upper limit on the age of the
SNR of $t \lesssim R/v_\mt{shock} \simeq 540\pm 50\;\mt{yr}$ (assuming
$v_\mt{shock} = 4/3\sigma_v$).

The first column in Table~\ref{pars} shows the best fit parameters for the
joint MOS+RGS analysis. However, we also fitted the RGS spectra separately
with a single NEI model. In this case the fitted parameters are quite
different, in particular the plasma temperature: $kT_e =
0.75^{+0.33}_{-0.09}$ keV and ionization timescale $n_et =
(1.28^{+0.25}_{-0.33})\times 10^{10}\;\mt{s/cm}^3$. The probable reason is
that the RGS is mostly sensitive to the energies in 0.5-1 keV, so it
misses the contribution from lines that mostly emit at higher temperatures
and the continuum. In particular the OVII lines around 0.55~keV/22~\AA\, has
much more statistical weight in the RGS spectra. The OVII line emission
probably has contributions from a lower temperature plasma, that is not
picked up by the MOS spectra. The discrepancy therefore points to the
presence of temperatures gradients in the SNR.

Ideally, one would like to incorporate those gradients into a more complete model
of the X-ray emission from SNR~0509-67.5. We made an attempt to obtain such
a model by fitting a joint MOS-RGS model with 3 NEI components, one for the
low temperature plasma, one for the bulk of the MOS spectrum, and one for
the Fe-K line emission. However, it turned out to be difficult to come up with
a unique, satisfying solutions, given the complexity of the parameter space.
Moreover, it would still involve some arbitrary choices. For example, it is
not clear what to do with the abundances; assume one set of abundances for
the first two NEI components, but then we do not allow for possible
abundance gradients that are also likely to be present. So in short, we have
reached here the limit of what is possible with multiple components. The RGS
data, however, do show that the models in Table~\ref{pars} are incomplete.

One of the advantages of the high spectral resolution of the RGS is
that it is possible to obtain an estimate of the nitrogen abundance.
Nitrogen is not an ejecta product of Type Ia supernovae, so any
nitrogen must come from the shocked CSM. 
Therefore, the nitrogen abundance, together with the emission measure
makes it possible to obtain an alternative estimate of the
pre-shock CSM density, avoiding some of the confusion due to mixture
of pure metal, ejecta components.

The hydrogen emission measure of the RGS fit is of $n_en_H\,V =
4\;\times 10^{59}\;\mt{cm}^{-3}$ and emitting volume of $V_X \simeq
10^{57}\;\mt{cm}^3$ yield $[n_en_H]^\mt{RGS} \simeq 400\;\mt{cm}^{-6}$. 
Taking into account best-fit RGS 
amount of nitrogen $X_N^\mt{RGS} = 0.01$ and typical
LMC abundance $X_N^\mt{CSM} = 0.4$ (both in solar units), we derive 
for the circumstellar matter $[n_en_H]^\mt{CSM} =
[n_en_H]^\mt{RGS}\,(X_N^\mt{RGS}/X_N^\mt{CSM})
= 10\;\mt{cm}^{-6}$. Thus for the shocked CSM (LMC abundances) $n_H =
3\;\mt{cm}^{-3}$, which yields $n_\mt{CSM} = 0.4 -
0.8\;\mt{cm}^{-3}$.

\FIGG rgs RGS_nei_wav RGS_2nei_vgau_A [width=0.8\hsize,angle=-90] 
Top: the RGS spectra with best-fit NEI model.
Bottom: the RGS spectra and the MOS+RGS best-fit NEI model.

\section{Numerical models} 
\label{models}
To simulate the evolution of the remnant and to compare it with the observed
data, we employed the hydrodynamical code {\sc supremna}, 
which is explained in details in \citet{sorokina}. 
The code assumes spherical symmetry, but it incorporates 
many relevant physical processes, without which it is not possible to
accurately predict peculiarities of SNR radiation, such as time-dependent
ionization, inner-shell collisional ionization, possible difference in
temperatures of electrons and ions, the influence of radiative losses, the
account of electron thermal conduction and nonthermal particles.
The calculations of the lines emission was based on
\citet{gaetzsalp} and \citet{mewe85}. The introduction of inner-shell
ionization processes is described in \citet{kosenko}.

To reproduce the observed remnant in numerical simulations we considered
various physical conditions of the CSM and also different explosion models:
deflagration --- W7 \citep{nomoto} and some of the `mildly-mixed''
delayed-detonation models \citep{woosley}.

Following the estimates of the CSM density derived from the {\sc spex}
spectral fitting, we surround the remnant with uniform medium of
$\rho_\mt{CSM} = 3\times10^{-25}\;\mt{g/cm}^3$. In this set up we found that
at the age of $t \sim 400$ years the modeled remnant's radius reaches $R
\sim 3.6$ pc, with typical velocities of the plasma of $v \sim 5000$ km/s.
These parameters are in agreement with the observed ones of the
SNR~0509-67.5. Modeled temperatures and X-ray spectra will be discussed in
details below.

\subsection{Basic modeling} 

First, we used a ``basic'' approach, described in~\citet{sorokina} where
electrons and ions are being heated partially due to artificial viscosity
and degree of the electrons heating is controlled by a parameter $q_i$, such
that $P_i = P_i(thermal) + q_iQ$ and $P_e = P_e(thermal) + (1-q_i)Q$, where
$P_e,\, P_i$ --- electrons and ions pressure, $Q$ --- artificial viscosity (is
the case when only collisional energy exchange takes place $q_i = 1 -
m_e/m_i$, $m_e,\,m_i$ -- electron and ion masses). Note that since $q_i$
controls the efficiency of electrons heating (their temperature may vary
theoretically from $10^{-4}T_i$ to $T_i$), this parameter has a strong
influence on the X-ray spectrum behavior (line fluxes with respect to
free-free continuum).

In this framework we have found that the best fit (in terms of
$\chi^2/d.o.f.$) for the observed spectrum gives the value of $q_i = 0.99$,
which yields the typical ratio of $T_e/T_i \simeq 3\times10^{-3}$ for the
case of W7 explosion model. We also considered a library of
thermonuclear explosion models by \citet{woosley}. We employed several of
them with various amounts of iron and intermediate mass elements (IME). We have found
that the explosion of a ``mildly-mixed'' delayed detonation model with
$E = 1.4\times10^{51}$ ergs, $M_\mt{Ni} = 0.5M_\odot, M_\mt{Fe} = 0.4M_\odot,
M_\mt{IME} = 0.3M_\odot$ 
\citep[c050403m model in][notation]{woosley}
gives the best-fit to the observed EPIC MOS spectrum of the SNR with $q_i =
0.9$ and thus $T_e/T_i \simeq 3\times10^{-2}$. These values seem to be in
reasonable agreement with the relation $T_e/T_i \propto v_S^{-2}$ ($v_S$ ---
shock velocity) proposed by~\citet{ghav06}. Indeed, assuming for the
SNR~0509-67.5 shock front speed of $v_S \simeq 4/3 v \simeq 6000$ km/s,
according to~\citet[][Fig.2]{ghav06} 
we should expect $T_e/T_i = 6\times10^{-3}$.

The simulated (for these two explosion models) and the observed X-ray
spectra are presented in \rfig{spqi}. Note that all of our models
exhibit noticeable lack of flux in the region of Fe~L complex (compared to
the observed spectrum). This is caused by some incompleteness of the atomic
data set for Fe~L emission employed in our package, as verified by directly
comparing a single NEI model as obtained by the {\sc supremna} code and in
{\sc spex}. This issue will be addressed in future updates of the {\sc
supremna} code.

At higher densities of the CSM, there is no the Fe~L emission deficit, but
in these models the expansion rate of the shell and the radius of the
contact discontinuity are in disagreement with the observed values
of the lines broadening and the remnant's size. So the value
of CSM density was chosen to satisfy the observed velocity of the emitting
ejecta, the age and the geometrical size (assuming that the observed edge of
the remnant close corresponds to the location of the contact discontinuity) of the
remnant for the considered explosion mechanisms. Figure~\ref{rhovr} shows
maximum velocity of the shocked ejecta and position of the contact
discontinuity depending on the assumed value of CSM density for the delayed
detonation explosion model c050403 at the age of 350 (left panels) and 400
(right panels) years. The figure indicates that for an assumed age of 400
years, the CSM density of $3\times10^{-25}\;\mt{g/cm}^3$ gives a reasonable
compromise between velocity and radius. An assumption that the remnant may
be as young as 350 years, allows the CSM to be denser than
$5\times10^{-25}\;\mt{g/cm}^3$, but in this case the synthetic X-ray spectra
disagree with the observed ones (there are major discrepancies in Fe~L
emission and Si~K and S~K lines fluxes). More energetic ($E >
1.4\times10^{51}$) progenitor could shift the allowable values of the
density to higher ranges. We checked several of them, which contain more
${}^56Ni$, and which could be consistent with the optical spectrum of the
supernova, as extracted from the light echo~\citep{rest08}. However, these
models contain too little Si and S to fit their emission lines.

\FIG rhovr rhovr_t [width=0.9\hsize,angle=0] 
Maximum velocity of the shocked ejecta ($v^{max}_{ej}$, upper row) and position of the
contact discontinuity ($R_{CD}$, lower row) depending on the value of CSM
density for 350 (left column) and 400 years (right column). Explosion model
--- delayed detonation c050403~\citep{woosley} with $E = 1.4\times10^{51}$
ergs. The observed values of the velocity and radius of the remnant are also
outlined.

X-ray spectra of the simulated and the observed (merged MOS1 and MOS2,
crosses) remnants. The theoretical spectra are based on W7 ($T_e/T_i \simeq
0.003$, $\chi^2/d.o.f. \simeq 27$, dashed line) and delayed-detonation
c050403m ($T_e/T_i \simeq 0.03$, $\chi^2/d.o.f. \simeq 19$, solid line)
explosion models, $\rhoc = 3\times10^{-25}\; \mathrm{g/cm}^3$, age $t = 400$
years.

The Table~\ref{hd_obs} shows a list of some basic properties of these
models. Some of the numerical values in the table are upper limits or just
typical for the corresponding case. Errors are presented where appropriate.

\FIG spqi W7c05_qi_sp1 [width=0.9\hsize,angle=0] X-ray spectra of the simulated and
the observed (merged MOS1 and MOS2, crosses) remnants. The theoretical
spectra are based on W7 ($T_e/T_i \simeq 0.003$, $\chi^2/d.o.f. \simeq 27$, dashed line) and
delayed-detonation c050403m ($T_e/T_i \simeq 0.03$, $\chi^2/d.o.f. \simeq
19$, solid line) explosion models. $\rhoc = 3\times10^{-25}\; \mathrm{g/cm}^3$, age $t = 400$
years.

\subsection{Accounting for the cosmic rays}

We introduced an additional component representing
the pressure and energy from (relativistic) cosmic ray particles,
in a way that could be easily implemented
within the current framework of this code. This was done by
altering (softening) the equation of state as follows:

\begin{equation}
\left\{
\begin{array}{rcl}
P &=& P_g + \frac 13 \hat a T^4 \\
E &=& E_g + \hat a T^4 \\
\end{array} 
\right.
\label{cr1}
\end{equation}
where $P_g$ is the gas pressure, $E_g$ is the gas internal energy, $T$ is
the ``temperature'' and $\hat a$ is a free parameter mimicking the
contribution of the relativistic particles. Effectively it means,
that some part of the energy is transferred into relativistic particles,
which behave similar to a photon gas.

Density profiles for the cases with and without this relativistic
correction for the delayed-detonation c050403m~\citep{woosley} model are
presented in~\rfig{c05kmrho}. Note that in the case of the relativistic
alteration of the equation of state (gray solid line), the density jump at
the forward shock is enhanced, compared to the basic set up. The density
profile of the ejecta behind the reverse shock became affected as well.

However, in reality the cosmic ray contribution to the internal
energy may be different for the forward and reverse shock. Recently,
\citet{heldervink} established, that in the case of the SNR Cas~A, the
reverse shock can accelerate cosmic ray particles up to energies of a
few TeV.

\FIG c05kmrho c05_3d24_400yr_r_d_CR [width=0.9\hsize,angle=0] 
Density profiles (solid lines) against radius for the hydrodynamical models
of age 400 years and for the delayed detonation explosion mechanism
c050403m. Upper panel: basic set up, lower panel: the set up with the
relativistic correction in the equation of state (Eq.~\ref{cr1}). Also the
distributions of Si (black dash-dotted line) and Fe (gray dashed line) in
the ejecta are shown.

In this set up, a configuration with the cosmic-ray energy density
$E_\mt{CR} = \hat a T^4$ of the order of $0.4E_{g}$, produce a model which
also gives a good fit to the observed spectrum. The modeled and observed
X-ray spectra for this approach are presented in \rfig{spcr}.

\citet{chev83} considered self-similar solutions for a mixture of a fluid
with $\gamma=5/3$ and  a relativistic gas (i.e., cosmic rays) with
$\gamma=4/3$. The effective adiabatic index was determined by a constant
fraction of cosmic rays in total pressure. Later \citet{blondinellison} used
a constant value of effective $\gamma$ which was taken sometimes even lower
than $4/3$ to allow for a stronger compression ratio mimicking radiative
shock waves. We note that our parametrization, being also simple in
comparison with a true cosmic rays model, is much richer than used by
\citet{chev83,blondinellison} in their purely hydrodynamic studies. We do
not assume a constant fraction of cosmic rays or a constant value of
effective $\gamma$, thermal conduction and radiative losses are fully
included into a hydrodynamical scheme, in contrast with the vast majority of
modern work on the subject. However, the validity of our simplified model of
cosmic ray contribution requires further tests and comparison with real
supernova remnants and more sophisticated schemes modeling them.

The value of the energy density found for relativistic particles $E_\mt{CR}$
can be used to estimate the intensity of magnetic field in the remnant. Assuming
that $E_\mt{CR}$ and $B^2/8\pi$ should be of the same order of magnitude, we
derive $B = 500 \pm 200\,\mu G$. This value appears to be somewhat high for
this SNR, but still within reasonable limits. For example, using the
correlation $\rho_\mt{CSM}v_S^3 \propto B^2$ from ~\citet[][Fig.~6]{jvink} 
one might expect $B \lesssim 100\,\mu G$.

The list of the typical properties of these models is also presented in
Table~\ref{hd_obs}. In this case the difference between electron and ion
temperatures is not so drastic: $T_e/T_i \simeq 0.03$, which is still in
agreement with \citet{ghav06}, where at $\gamma = 4/3$ (relativistic
equation of state) the SNR plasma velocities (of $\sim 4500$ km/s)
correspond to $T_e/T_i \simeq 10^{-2}$.

\begin{table*} 

\caption{Observed properties of the SNR~0509-67.5 and the properties of the numerical
models: the one with the ``basic'' set up and one with a relativistic component (Eq.\ref{cr1}). The
table columns are: R -- visible remnant's radius (in the models
this is the position of the contact discontinuity), t -- the age,
$n_\mt{CSM}$ -- circumstellar number density, $v$ -- velocity of the shocked
ejecta (maximum value for the numerical models), $kT_e$ -- electron
temperature (for numerical models it is the maximum temperature), $kT_i$ --
maximum ion temperature, $M_\mt{Fe}$ -- an estimate of the swept up iron
mass (in solar masses), $N_H$ -- column density on the line of sight
\citep[derived from {\sc xspec},][spectral fitting]{xspec}.}

\begin{center}
\begin{tabular}{lllllllll}
\hline \\  Parameters  & R, pc & t, yrs & $n_\mt{CSM}\; \mt{cm}^{-3}$ & $v$, km/s & $kT_e$, keV & 
$kT_i$, keV &  $M_\mt{Fe}, M_\odot$ & $N_H\;10^{21}\mt{cm}^{-2}$ \\
\hline\hline \\  SNR 0509-67.5 & 3.6 & $\lesssim 500$ & $ 0.4 - 0.6$ &  $4900\pm 400$   & 
$2.5 - 3.6$ & - & $\sim 0.05$  & -         \\
\hline\hline \\  W7(basic) & 3.6 & $400$         &  $0.1$          & $\lesssim 4600$          & 
$0.5 - 3.6$    & $ 150 - 2500 $  & $\sim 0.12$ & $0.7\pm 0.3$ \\
\hline \\  c050403m(basic) & 3.7 & $400$   &  $0.1$        &  $\lesssim 4200$          & 
$1 - 1.5$    & $30 - 40 $  & $\sim 0.36$ & $5.6\pm 0.3$ \\
\hline\hline \\  W7(CR) & 3.6 & $400$         &  $0.1$          & $\lesssim 4300$          & 
$1.8 - 1.9$    & $ 20 - 36 $  & $\sim 0.12$ & $2.1\pm 0.3$ \\
\hline \\  c050403m(CR) & 3.8 & $400$   &  $0.1$        &  $\lesssim 4700$          & 
$2 - 45$    & $ 30 - 300 $  & $\sim 0.36$ & $2.6\pm 0.3$ \\
\hline
\end{tabular}
\end{center}
\label{hd_obs}
\end{table*}

\FIG spcr W7c05_CR_sp1 [width=0.9\hsize,angle=0] X-ray spectra of the
simulated and the observed (merged MOS1 and MOS1) remnants.  The theoretical
spectra are based on W7 ($\chi^2/d.o.f. \simeq 21$, dashed line) and
delayed-detonation c050403m ($\chi^2/d.o.f. \simeq 13$, solid line)
explosion models. $\rhoc = 3\times10^{-25}\; \mathrm{g/cm}^3$, age $t = 400$
years, $E_\mt{CR}/E_{g} \simeq 0.4$.

\section{Discussions and concluding remarks} 
\label{discs}
The analysis of the XMM-Newton X-ray spectrum of the SNR~0509-67.5 allowed
us to estimate basic parameters of the object and reveals several peculiar
features.

First of all we obtained reasonable and consistent estimates for the CSM
density, using various methods: the EPIC MOS data fitting gives the
estimates of the circumstellar medium density of $n_\mt{CSM} \lesssim
0.6\;\mt{cm}^{-3}$, the RGS spectral fitting of nitrogen abundance (which is
not a product of thermonuclear explosion) yield $n_\mt{CSM} =
0.4-0.8\;\mt{cm}^{-3}$. In the hydrodynamical simulations we have
found similar, but somewhat lower value of $n_\mt{CSM} = 0.1\;\mt{cm}^{-3}$.

The abundances deduced from a single ionization timescale NEI  (\rfig{abs})
show that swept up amount of iron in the remnant is low. Moreover the model
was unable to reproduce the observed Fe~K line, so the line was fitted
separately. The centroid of Fe~K line is located at $6.50\pm 0.05$~keV, so it
should be produced by low ionized ($\leqslant$ XVII) ions of iron. This
could mean that the reverse shock of the remnant just recently reached and
heated up the iron core of the supernova. The Fe~K feature enables us to
make a crude estimate of the amount of heated iron of $M_\mt{Fe} \sim
0.1\,M_\odot$.

In a framework of a NEI model with fixed single ionization timescale,
abundances of the species lighter than aluminum tend to be overestimated,
meanwhile, the amount of heavy species (iron group) tends to be underestimated
\citep{LMC_hughes}. For example, in the W7 model the Mg/Si line flux ratio
is higher than  observed (\rfig{spqi}, \rfig{spcr}), while~\rfig{abs} shows
that the Mg/Si abundance ratio is lower compared to the observed one. The
same is true for the iron abundance: \citet{badenes08, rest08} showed that
the remnant has a very large amount of iron, while the abundance of Fe,
deduced from this simple fitting is considerably low. Note, however, that
probably most of the Fe has not yet been heated by the reverse shock.

In this context (using only abundances derived from single ionization
timescale NEI model) it is hard to determine the precise explosion
mechanism. If we consider only the species from Si to Ar, then a
delayed-detonation model with $M_\mt{Fe} = 0.9\,M_\odot$ and an explosion
energy of $1.4\times10^{51}$ ergs \citep{woosley} seems more preferable
(\rfig{abs}). The same tendency is confirmed by comparison of the
observed spectra and the spectra from numerical simulations, performed in
this study. This conclusion is in agreement with the results obtained by
\citet{badenes08,rest08}.

A separate study of the remnant's bright southwest region shows that this
spot appears to be more metal-rich: our fitting shows that it is at
least twice more abundant in metals than the average metallicity throughout the
remnant. This points to either an asymmetric explosion of the SNR or asymmetric
CSM. Meanwhile note, that in \citet{parviz_uv} it was mentioned that this
region has a blueshifted velocity excess. Which is a hint for an
asymmetric explosion.

The numerical models are in agreement with the overall properties of
SNR~0509-67.5 inferred from the studies presented in Section~\ref{spectra},
which are as following: the CSM density $\rho_\mt{CSM} \simeq
3\times10^{-25}\;\mt{g/cm}^3$, the age $t = 350 - 400$ years, and the plasma
velocities $v \simeq 4000 - 5000$ km/s. The ratio of electron to ion
temperatures of $T_e/T_i \simeq 0.01$ is in satisfactory agreement with
the expansion velocity of the remnant~\citep{ghav06}, but heavily depends on
the assumed explosion model \citep[see also][]{badenes}. Ejecta abundances
govern electron population $n_e$ and thus, the X-ray spectral shape. The
latter is used to find the ``best-fit'' value of $T_e/T_i$.

The low value of $T_e/T_i$ together with almost undetectable power-law
emission point to the yet ineffective role of plasma instabilities and/or
magnetic field at this stage in this object. Nevertheless, the simulations
also indicate that the energy of relativistic particles is probably not
negligible and may contribute up to $\sim 40\%$ of thermal energy, so the
magnetic field (provided that $E_\mt{CR} \sim B^2/8\pi$) might be slightly
higher, $\sim 500\,\mu G$, than the expected value of $\sim 100\,\mu G$
based on the relation reported by \citet{jvink}, but within reasonable
limits.

We have found satisfactory hydrodynamical models for the remnant, but many
problems still exist. 

In our synthetic spectra some of the line centroids and line fluxes ratios
do not perfectly fit the observations. This is partly due to limited set of
explosion models available and to still appreciable number of free
parameters (electron temperature, thermal conduction, magnetic fields,
cosmic ray particles, and other non-thermal and relativistic effects). For
instance, more energetic explosion model ($E > 1.4\times10^{51}$) could
produce a supernova remnant at the age of $350 - 400$ years with required
high ejecta velocity, observed radius, and more acceptable higher value of
CSM density (up to $10^{-24}\;\mt{g/cm}^3$). Nevertheless such a model
should contain a lot of ${}^{56}$Ni and small amount of Si and S. The X-ray
spectra produced by such an explosion model disagree with the observed one
which contains a very well pronounced Si~K and S~K lines.

Moreover, one of the shortcomings of the code is that for the
calculation of the X-ray emission, and for the reproduction of the observed
spectrum around 1 keV, we need to update the atomic data for Fe~L code. 
A task that we plan to undertake in the near future.

To create more or less reliable models for supernova remnants, further
developments of the physics in hydrodynamical simulations are required.
Self-consistent treatment of cosmic-rays acceleration is necessary. 3D
simulations may shed some light on the influence of various instabilities on
the SNR dynamics. Further analysis of high resolution X-ray spectra is
needed to evaluate the conditions in other SNRs as well.

\begin{acknowledgements}
We are grateful to the referee C.~Badenes for valuable comments 
which helped to improve the paper. We also thank P.~Lundqvist for providing
his X-ray code which is incorporated in the package {\sc supremna},
J.~Kaastra for helpful advice on X-ray line emission issues, 
M.~Gilfanov for valuable pointers to X-ray data sources and S.~Woosley for
his set of SNIa models used in our work.

DK and JV are supported by a Vidi grant from the Netherlands Organization for
Scientific Research (PI J.~Vink), The work of DK is also partially supported
in Russia by RFBR under grants 05-02-17480, 06-02-16025, 07-02-00961, and by Russian
Leading Scientific School Foundation under grant RLSS-2977.2008.2.

SB is supported in Russia partly by grants RFBR 07-02-00830-a, 
RLSS-3884.2008.2 and by grant IB7320-110996/1 of the Swiss National Science
Foundation.
\end{acknowledgements}

\bibliographystyle{aa}
\bibliography{kosenko.bib}

\end{document}

%% file: defs.tex
\usepackage{graphicx}
\usepackage{graphpap}
\usepackage{epsf}
\usepackage{amssymb}
\usepackage{color}

\usepackage{natbib}

\evensidemargin=\oddsidemargin \textheight=23cm \textwidth=16.5cm

\def\FIG #1 #2 [#3] #4\par{%
  \begin{figure}\begin{center}%
    \includegraphics*[#3]{#2}%
    \\
    \caption{#4}%
    \label{#1}%
  \end{center}\end{figure}%
}
\def\FIGG #1 #2 #3 [#4] #5\par{%
  \begin{figure}
               \includegraphics*[#4]{#2}
               \includegraphics*[#4]{#3}
    \caption{#5}%
    \label{#1}%
  \end{figure}%
}

\def\FIGs #1 #2 #3 #4 #5 [#6] #7\par{%
  \begin{figure}[ht]\begin{center}%
        \includegraphics[#6]{#2}
        \includegraphics[#6]{#3} \\
        \includegraphics[#6]{#4}
        \includegraphics[#6]{#5}
        \caption{#7}
        \label{#1}
   \end{center}\end{figure}
}
\def\FIGss #1 #2 #3 #4 #5 #6 #7 [#8] #9\par{%
  \begin{figure}[ht]\begin{center}%
        \includegraphics*[#8]{#2}
        \includegraphics*[#8]{#3}\\
        \includegraphics*[#8]{#4}
        \includegraphics*[#8]{#5}\\
        \includegraphics*[#8]{#6}
        \includegraphics*[#8]{#7}
        \caption{#9}
        \label{#1}
    \end{center}\end{figure}
}

\def\rfig#1{Fig.\ref{#1}}

\def\rhoc{\rho_\mathrm{CSM}}

\newcommand{\mt}[1]{\mathrm{#1}}